\documentclass[aps,pre,twocolumn,groupedaddress]{revtex4}
\usepackage{bm}
\usepackage{color}
\usepackage{amsmath}
\usepackage{graphicx}
\usepackage{amssymb}

\begin{document}

\title{Dynamics and polarization of superparamagnetic chiral nanomotors in a rotating magnetic field}

\author{Konstantin I. Morozov$^1$}
\author{Alexander M. Leshansky$^{1,2}$}
\email{lisha@tx.technion.ac.il}
\affiliation{$^1$Department of Chemical Engineering, Technion -- Israel Institute of Technology, Haifa 32000, Israel}
\affiliation{$^2$Technion Autonomous System Program (TASP), Haifa 32000, Israel}

\begin{abstract}
Externally powered magnetic nanomotors are of particular interest due to the potential use for \emph{in vivo} biomedical applications. Here we develop a theory for dynamics and polarization of recently fabricated superparamagnetic chiral nanomotors powered by a rotating magnetic field. We study in detail various experimentally observed regimes of the nanomotor dynamic orientation and propulsion and establish the dependence of these properties on polarization and geometry of the propellers. Based on the proposed theory we introduce a novel ``steerability" parameter $\gamma$ that can be used to rank polarizable nanomotors by their propulsive capability. The theoretical predictions of the nanomotor orientation and propulsion speed are in excellent agreement with available experimental results. Lastly, we apply slender-body approximation to estimate polarization anisotropy and orientation of the easy-axis of superparamagnetic helical propellers.
\end{abstract}

\maketitle
\date{\today}

\section*{Introduction}

The emergent interest in artificial "nature-inspired" micro- and nano-structures that can be remotely actuated, navigated and delivered to a specific location \emph{in-vivo} is largely driven by the immense potential this technology offers to biomedical applications.  Several approaches are currently of interest ranging from catalytically driven (chemical-fuel-driven) nanomotors 
to thermally, light and ultrasound-driven colloids (see \cite{wang} for state-of-the-art review of the subject).  An alternative approach relies on externally powered nanomotors, where the particle is propelled through media by an external magnetic field. This allows contact-free and fuel-free propulsion in biologically active systems without chemical modification of the environment. In particular, it was shown \cite{GF,N2,N3}  that a weak \emph{rotating} magnetic field can be used efficiently to propel \emph{chiral ferromagnetic nanomotors}. These nanohelices are magnetized by a strong magnetic field and retain remanent magnetization when stirred by a relatively weak (of the order of a few milli Tesla) rotating uniform magnetic field. The typical propulsion speeds offered by this technique are four-five orders of magnitude higher than these offered by the traditional techniques based on gradient of magnetic field, e.g. \cite{Lubbe}. In the past few years the new technique has attracted considerable attention \cite{Tot,G,Peyer,G_new}. Various methods, such as ``top-down" approach \cite{N2}, glancing angle deposition \cite{GF} and direct laser writing (DLW) combined with vapor deposition \cite{Tot}, have been developed toward fabrication of ferromagnetic nanohelices. Experiments \cite{G,G_new} showed that at low frequency of the rotating magnetic field nanomotors \emph{tumble} in the plane of the field rotation without propulsion. However, upon increasing the field frequency, the tumbling switches to \emph{wobbling} and the precession angle (between the axis of the field rotation and the helical axis) gradually diminishes and a corkscrew-like \emph{propulsion} takes over.

Most recently an alternative method for microfabrication of \emph{superparamagnetic} nanomotors was reported \cite{Suter0,Suter1}. The method relies on DLW and two-photon polymerization of a curable superparamagnetic polymer composite. These helices do not possess remanent magnetization, but are magnetized by the applied magnetic field. The advantages of using superpamagnetic polymer composite are the ease of microfabrication as magnetic material is already incorporated into the polymer (no need for thin film deposition), low toxicity \cite{Suter1} and the lack of magnetically-driven agglomeration of adjacent nanomotors in the absence of an applied field. Qualitatively, the dynamics of superparamagnetic helices resembles that of ferromagnetic nanomotors, i.e. they exhibit tumbling, wobbling and propulsion upon increasing frequency of the driving field. The magnetic properties of superparamagnetic nanohelices are characterized by their effective magnetic susceptibility and orientation of the magnetic \emph{easy-axis}, generally dominated by the geometric effects \cite{Nelson}. Orientation of the easy-axis plays a central role in controlling the dynamics of superparamagnetic nanomotors in a way similar to orientation of the magnetic moment of permanently magnetized nanohelices. It was recently shown that orientation of the magnetic easy-axis in  polymer composites can be manipulated in order to minimize tumbling/wobbling and maximize propulsion, by aligning (otherwise randomly dispersed) superparamagnetic nanoparticles prior to crosslinking the polymer matrix  \cite{Peters}.

Physically, the dynamics of the magnetic nanomotors is governed by the interplay of magnetic and viscous forces. Despite the high interest, the theory for the dynamics of magnetically driven nanomotors is quite limited. A recent study \cite{KWS13} addressed optimization of the chirality that maximizes the propulsion speed at a prescribed driving field assuming perfect alignment of the helix along the axis of the field rotation. In Refs. \cite{ML13,G12} the hydrodynamic aspects of wobbling-to-swimming transition for a helix with purely transverse permanent magnetization were studied asymptotically and numerically showing a qualitative agreement with experiments. Ghosh et al. \cite{G_new} found the formal mathematical solution for the orientation of permanently magnetized nanomotors. In \cite{ML} we studied in detail the dynamics of ferromagnetic chiral nanomotors and established the relation between their orientation and propulsion with the actuation frequency, remanent magnetization and the geometry. The theoretical predictions for the transition threshold between regimes and nanomotor alignment and propulsion speed in \cite{ML} showed excellent agreement with available experimental results.

No theoretical study concerning the dynamics of polarizable superparamagnetic nanomotors (e.g. in \cite{Peters,Nelson,Suter0,Suter1}) has been reported. In the present paper we propose an original theory for dynamics and polarization of superparamagnetic helical propellers and compare the predictions of the theory to previously reported experimental results.

\section*{Polarizable helix in rotating magnetic field: problem formulation}

\begin{figure}[h] \centering
\includegraphics[width=0.3\textwidth]{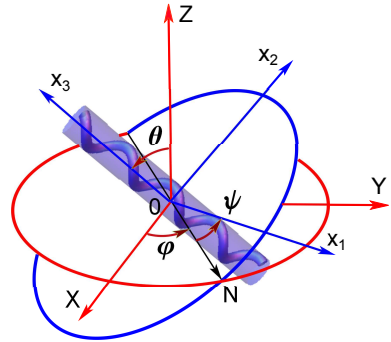}
\caption{
The laboratory and body-fixed coordinate systems with the corresponding axes
$XYZ$ and $x_1x_2x_3$ and the definition of the Euler
angles $\varphi$, $\theta$ and $\psi$. For simplicity, the cylindrical
envelope of a helix is shown.
\label{fig:1}}
\end{figure}

Let us consider the dynamics of polarizable helix in the external rotating
magnetic field. We use two different coordinate frames --  the laboratory coordinate system (LCS)
fixed in space and the body-fixed coordinate system (BCS) attached
to the helix (see Fig.~\ref{fig:1}). The coordinate axes of the two frames are
$XYZ$ and $x_1x_2x_3$, respectively. We denote by $\bm{H}$ the externally imposed
rotating (uniform) magnetic field. We also assume that in the LCS the field rotates in the $XY$-plane
\begin{equation}
\bm{H}^{LCS}=H(\cos \omega t, \sin \omega t, 0) \,, \label{eq:field}
\end{equation}
where $H$ and $\omega$ are, correspondingly, the field amplitude and angular frequency.

Once the external field (\ref{eq:field}) is turned on, the helix polarizes.
Owing to the magnetic anisotropy of the helix, the polarization
vector ${\bm M}$ is not generally aligned with $\bm{H}$. We assume uniaxial magnetic anisotropy of the helix with director ${\bf n} $. The
general form of the uniaxial magnetic susceptibility tensor ${\boldsymbol {\chi}}$ is
\cite{Gennes}
\begin{equation}
 \chi _{ik}=\chi_0 \delta_{ik}+\textstyle \Delta \chi (n_in_k-\frac 13 \delta_{ik}) \,, \label{eq:chi}
\end{equation}
where $\chi_0$ is the isotropic part of the magnetic susceptibility, $\delta_{ik}$ is the delta-symbol,
$\Delta \chi = \chi_\|-\chi_\perp$ is the scalar parameter of magnetic anisotropy with $\chi_\|$ and $\chi_\perp$
being the main components of the tensor ${\boldsymbol {\chi}}$ along the anisotropy axis and in the transverse direction,
respectively. In this paper we assume \emph{easy-axis anisotropy}, i.e.,
the positive values of parameter $\Delta \chi$. As we will see below, the case of an \emph{easy-plane anisotropy} characterized by
the negative values of $\Delta\chi$ is less relevant towards the present study since the helix cannot propel.
We assume an arbitrary angle $\Phi$ between the director ${\bm n}$ and the helical $x_3$-axis. In the BCS  ${\bm n}$ can be written in the form
\begin{equation}
{\bm n}^{BCS}=(\sin \Phi, 0, \cos \Phi) \,, \label{eq:n}
\end{equation}
The orientation of the BCS with respect to the LCS is determined by three
Euler angles $\varphi$, $\theta$ and $\psi$ \cite{LL}, as shown schematically in Fig.~\ref{fig:1}.

The rotational magnetic field (\ref{eq:field}) polarizes the helix producing the magnetic moment ${\boldsymbol {\mathcal {M}}}={\bm M}V={\boldsymbol {\chi}}\cdot{\bm H}V$ (with $V$ being the helix volume), and also the magnetic torque $\bm{L}_m={\boldsymbol {\mathcal {M}}}\times\bm{H}$. Substituting the expression for susceptibility, we find that
\begin{equation}
\bm{L}_m=\Delta \chi H^2V [{\bm n}\times \bm{h}]({\bm n}\cdot \bm{h})\,, \label{eq:Lm}
\end{equation}
where $\bm{h}=\bm{H}/H$ is the unit vector of the external field.

This torque is a source of both rotational and translational movements of the particle. In the Stokes approximation, the helix motion is governed by the balance of external and viscous forces and torques
acting on the particle \cite{HB}
\begin{eqnarray}
0= {\boldsymbol {\xi}}\cdot{\bm U}+ {\boldsymbol {\mathcal {B}}}\cdot \bm{\mathit{\Omega}}\,, \label{eq:u1}\\
\bm{L}_m={\boldsymbol {\mathcal {B}}}^T\cdot{\bm U}+ {\boldsymbol {\kappa}}\cdot\bm{\mathit{\Omega}}\,, \label{eq:u2}
\end{eqnarray}
where $\bm{U}$ and $\bm{\mathit{\Omega}}$ are the translational and angular velocities of helix, ${\boldsymbol {\xi}}$, ${\boldsymbol {\kappa}}$ and ${\boldsymbol {\mathcal {B}}}$ are the translation, rotation and coupling viscous resistance tensors, respectively \cite{HB}. We have also assumed in Eq.~(\ref{eq:u1}) that no external force is exerted on the helix.

The formal solution of the problem can be readily obtained from Eqs.~(\ref{eq:u1}), (\ref{eq:u2}):
\begin{equation}
{\bm U}=-{\boldsymbol {\xi}}^{-1}\cdot{\boldsymbol {\mathcal {B}}}\cdot \bm{\mathit{\Omega}}\,,\,\,\,\bm{\mathit{\Omega}}={\boldsymbol {\kappa}_\mathrm{eff}^{-1}\cdot\bm{L}_m}\,, \label{eq:u3}
\end{equation}
where $ {\boldsymbol {\kappa}}_\mathrm{eff}= {\boldsymbol {\kappa}}- {\boldsymbol {\mathcal {B}}}^T \cdot{\boldsymbol {\xi}}^{-1}\cdot {\boldsymbol {\mathcal {B}}}$ is the re-normalized viscous
rotation tensor.

The problem of the helix dynamics can be decomposed into two separate problems: (i) rotational
motion of an \emph{achiral} slender particle i.e. with ${\boldsymbol {\mathcal {B}}}=0$ and diagonal $\bm \kappa$ with components $\kappa_{11}=\kappa_{22}=\kappa_\perp>\kappa_{33}=\kappa_\|$, e.g. axially symmetric slender particle, such as cylinder or prolate spheroid enclosing the helix (see Fig.~\ref{fig:1}), and (ii) translation of a \emph{chiral} particle rotating with a prescribed angular velocity found in (i) (see \cite{ML} for detailed justification of such decomposition.

In the following sections we consider both problems.

\section*{Polarizable cylinder in a rotating magnetic field}

It is convenient to right down the equation of the rotational motion (the second equation in Eqs.~(\ref{eq:u3}))
in the BCS in which the tensor ${\boldsymbol{\kappa}}$ takes a diagonal form \cite{note1}.
The director ${\bm h}$ of the magnetic field (\ref{eq:field}) in the
BCS is $ \bm{h}^{BCS}=\bm {R}\cdot\bm{h}$, where $\bm {R}$ is the rotation matrix \cite{D} (see Appendix A).
Substituting components of the angular velocity $\bm{\mathit{\Omega}}$ \cite{LL} into the second equation in (\ref{eq:u3}), the torque balance takes the form:
\begin{eqnarray}
A({\bm n}\cdot \bm {R} \cdot \bm{h})[{\bm n}\times (\bm {R} \cdot \bm{h})]_{x_1} =\dot{\varphi}s_{\theta}s_{\psi}+\dot{\theta}c_{\psi}\,, \label{eq:1}\\
A({\bm n}\cdot \bm {R} \cdot \bm{h})[{\bm n}\times (\bm {R} \cdot \bm{h})]_{x_2}=\dot{\varphi}s_{\theta}c_{\psi}-\dot{\theta}s_{\psi}\,, \label{eq:2}\\
pA({\bm n}\cdot \bm {R} \cdot \bm{h})[{\bm n}\times (\bm {R} \cdot \bm{h})]_{x_3}=\dot{\varphi}c_{\theta}+\dot{\psi}\,. \label{eq:3}
\end{eqnarray}
Here $A=\Delta \chi H^2V/\kappa_\perp$ is the characteristic frequency of the problem.
We also use here the compact notation throughout the paper, i.e. $s_{\psi}=\sin{\psi}$, $c_{\theta}=\cos{\theta}$, etc. and the dot stands for the time derivative.
The rotational friction coefficient ratio $p=\kappa_\perp/\kappa_\| \gtrsim 1$ depends on the aspect ratio of the cylinder: it is $p\simeq 1$ for a short cylinder (i.e. disk), for which $\kappa_\|\simeq \kappa_\perp$ and increases with the growth of the aspect ratio \cite{HB}.

Generally,  overdamped dynamics of a magnetic particle in a rotating magnetic field
can be realized via \emph{synchronous} and \emph{asynchronous} regimes \cite{Pincus,Cebers1}.
The synchronous regime is observed when there is a constant phase-lag between the Euler angle $\varphi$
of the particle body and the external magnetic field $\bm{H}$ while the angles $\theta$
and $\psi$ are not varying with time. As we show in the next section, the solution to the problem in
the synchronous regime can be found analytically.

\subsection*{Synchronous regime: low-frequency tumbling solution}

The low-frequency tumbling solution can be obtained by using the following ansatz for the Euler angles: $\psi=0$,
$\theta=\pi/2$,  $\varphi=\omega t-\varphi_0$, where $\varphi_0$ is a constant.
With these values, the components of unit vector $\bm {h}$ in the BCS become
\begin{equation}
(\bm {R} \cdot \bm{h})_{x_1}=c_{\varphi_0}\,,\,\,(\bm {R} \cdot \bm{h})_{x_2}=0\,,\,\,(\bm {R} \cdot \bm{h})_{x_3}=-s_{\varphi_0}\,. \label{eq:h1}
\end{equation}
As a result, Eqs.~(\ref{eq:1}) and (\ref{eq:3}) are
satisfied identically, whereas Eq.~(\ref{eq:2}) determines the constant $\varphi_0$:
\begin{equation}
As_{2(\Phi-\varphi_0)}=2\omega\,. \label{eq:8}
\end{equation}
\begin{figure}[h] \centering
\includegraphics[width=0.20\textwidth]{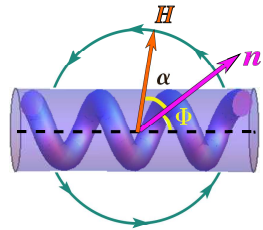}
\caption{
The low-frequency synchronous regime. Both the helical axes and axes of magnetic anisotropy
lie in the field rotation plane. The anisotropy axes $\bm n$ inclined at an angle $\Phi$
to the helical axis. The rotating magnetic field outruns $\bm n$ by the angle $\alpha$.
\label{fig:lowfreq}}
\end{figure}

The solution describes a \emph{tumbling} regime, e.g., helix rotation about its short axis.
The main feature of this regime is the fact
that the easy-axis $\bm n$ and the $x_3$-axis of the cylinder both lie
\emph{in the field plane}, as shown in Fig.~\ref{fig:lowfreq}.
Physically, this regime can be understood
from the next qualitative argument. In the constant magnetic
field, the helix magnetic moment ${\boldsymbol {\mathcal {M}}}$
and easy-axis $\bm n$  are oriented along the field
direction; the orientation of the $x_3$-axis is arbitrary, while it forms a solid angle $\Phi$ with $\bm n$.
In the weakly (quasi-statically) rotating magnetic field,
${\boldsymbol {\mathcal {M}}}$ and
$\bm n$ rotate with a small lag behind the external field $\bm H$. Thus,
vectors ${\boldsymbol {\mathcal {M}}}$ and
$\bm n$ both prove to be in the plane of the field rotation. The degeneracy
in the orientation of the propeller, however, is removed:
the viscous friction causes the $x_3$ axis to align in the plane of the field rotation.
The difference $\alpha=\Phi-\varphi_0$ in Eq.~(\ref{eq:8}) determines
the outrunning  angle of the rotating field $\bm H$ relatively
to the easy-axis $\bm n$ (see Fig.~\ref{fig:lowfreq}).
In the  static magnetic field, $\omega=0$, and both vectors coincide, i.e. $\alpha=0$.
As seen from Eq.~(\ref{eq:8}), the
solution exists within the limiting interval of field frequencies, from $\omega=0$, up to the
maximal value $\omega_{\textrm I}=A/2$.
When $\omega=\omega_{\textrm I}$, the angle $\alpha=45^\circ$
and the magnetic torque $\bm{L}_m$ reaches its maximal value (see Eq.~(\ref{eq:Lm}) and left-hand side of Eqs.~(\ref{eq:1}) and (\ref{eq:8})),
i.e., a further increase of the field frequency leads to the breakdown of the synchronous rotation and transition to the asynchronous regime.

There is, however, an additional synchronous solution that branches from the tumbling solution one at the \emph{finite value} of
the driving frequency $\omega_{*}$ prior to transition to the asynchronous regime.
The transition to this additional \emph{high-frequency (wobbling) solution} can be expected by
the following reasoning applicable for \emph{slender (rod-like) particles}.
The low-frequency tumbling solution illustrated in Fig.~\ref{fig:lowfreq}
is characterized by high viscous friction
owing to the propeller rotation about its short axes. The rotation around
the longer $x_3$-axis would be accompanied by a significant
reduction of the viscous friction, but at the same time, by higher value of magnetic energy $E_m=-({\bm H}\cdot {\boldsymbol {\chi}} \cdot \bm{H})V/2$ \cite{LL8}.
Therefore, there is a \emph{competition} between the magnetic and frictional torques. In the low-frequency tumbling regime, the viscous friction is of secondary importance -- it is only responsible for removing the degeneracy of the propeller's orientation, as any deviations from $\theta=0$ (i.e. precession) are suppressed by the viscous torques acting to bring the $x_3$-axis
back to the plane of field rotation.

Upon increasing the driving frequency, $\omega$, however, the role of the
viscous forces increases and their interplay with the magnetic forces results into the new high-frequency
wobbling regime.

Concluding this section, we point out that the higher the particle slenderness, the more pronounced is
the competition between the magnetic and viscous forces. In contrast, in the case of disk-like platelets
(see, e.g., Ref.~\cite{Erb}), the anisotropy of rotational friction coefficient is negligible,
$p \approx 1$, and orientation of such platelets is determined solely by the magnetic forces.
As a result, platelets rotate in such a way that the plane formed by their two major
eigenvectors of the susceptibility tensor align with the plane of the rotating magnetic field
for all field frequencies.
There are also two potential cases with polarizable helices (with $p>1$) where the competition between the magnetic and viscous
forces cancels out. (i) The case of positive magnetic anisotropy, $\Delta \chi>0$, when easy-axis $\bm n$ is strictly perpendicular to the
helix axis $x_3$. This polarization is optimal for the helix propulsion: the magnetic
field enforces the helix to spin around its longer axis with minimal rotational friction, i.e. the tumbling is prevented for all driving frequencies. The recently fabricated superparamagnetic micro-helices with adjusted magnetic anisotropy in \cite{Peters} possess this type of polarization. (ii) For the case of the easy-plane anisotropy, $\Delta \chi<0$, the rotating magnetic field drives the helix to align its easy-plane with
the plane of the field rotation. However, since both principal polarization axes in the easy-plane are equivalent,
the lack of magnetic anisotropy in the field plane would yield no corkscrew-like rotation and propulsion.

\subsection*{Synchronous regime: High-frequency wobbling solution}

The high-frequency wobbling solution can be found by using the following ansatz for the $\varphi$-Euler angle:
${\varphi}=\omega t$.
Thus the projections of the unit vector $\bm {h}$ in the BCS are
\begin{equation}
(\bm {R} \cdot \bm{h})_{x_1}=c_{\psi}\,,\,\,(\bm {R} \cdot \bm{h})_{x_2}=-s_{\psi}\,,\,\,(\bm {R} \cdot \bm{h})_{x_3}=0\,. \label{eq:h2}
\end{equation}
The system of three equations (\ref{eq:1})-(\ref{eq:3}) reduces to the
following system of two equations governing the two
remaining Euler angles, $\psi$ and $\theta$:
\begin{eqnarray}
\omega_{*}c_{\psi}={\omega}s_{\theta}\,, \label{eq:sol21} \\
-\gamma \omega_{*}s_{2\psi}=2{\omega}c_{\theta}\,, \label{eq:sol22}
\end{eqnarray}
where we introduced the critical frequency $\omega_{*}=(A/2)s_{2\Phi}$
and  the ``\emph{steerability}'' parameter $\gamma=p\tan \Phi$ (see the Discussion Sec. for details). There is a non-trivial solution of (\ref{eq:sol21}--\ref{eq:sol22}) provided that $\gamma=p\tan \Phi \geq 1$,
\begin{equation}
c_{\psi}= \frac {(1+\gamma^2)^{1/2}}{\gamma \sqrt{2}} \left(1+ {\sqrt{1-\omega^2/\omega_{\textrm {II}}^2}}\right)^{1/2}\,, \label{eq:sol23}
\end{equation}
\begin{equation}
s_{\theta}= \frac {\sqrt{2}}{(1+\gamma^2)^{1/2}}\frac {\omega_{\textrm {II}}}{\omega} \left(1+ {\sqrt{1-\omega^2/\omega_{\textrm {II}}^2}}\right)^{1/2}\,. \label{eq:sol24}
\end{equation}
Here we used the notation
\begin{equation}
\omega_{\textrm {II}}=\omega_{*}\frac{1+\gamma^2}{2\gamma}=\frac {\Delta \chi H^2 V}{{\kappa_\perp}} \frac {1+\gamma^2}{4\gamma}\sin 2\Phi\,.\label{eq:so}
\end{equation}
The high-frequency solution (\ref{eq:sol23}--\ref{eq:sol24})
branches from the low-frequency solution (\ref{eq:8}) at frequency
$\omega=\omega_{*}$ where $\theta=\pi/2$
and persists in the limiting frequency interval $\omega\in [\omega_{*}, \omega_{\textrm {II}}]$.

The high-frequency solution corresponds to \emph{wobbling} dynamics where the
increase in driving frequency from $\omega_{*}$ up to $\omega_{\textrm {II}}$ yields the gradual decrease in
the angle $\theta$ between the propeller's $x_3$-axis and the $Z$-axis of the field rotation, or the precession angle.
The minimal precession angle $\theta_\mathrm{min}$ is attained at $\omega=\omega_\mathrm{II}$
\begin{equation}
(s_{\theta})_{min}=\frac {\sqrt{2}}{(1+\gamma^2)^{1/2}}\,. \label{eq:angle4}
\end{equation}
The maximal frequency, $\omega_\mathrm{II}$, is usually termed as \emph{step-out frequency} $\omega_{s-o}\equiv\omega_\mathrm{II}$. At frequencies $\omega>\omega_{s-o}$ the high-frequency solution breaks down and the synchronous regime switches to the asynchronous one.

Finally, note that the high-frequency solution (\ref{eq:sol23}--\ref{eq:sol24}) requires $\gamma>1$, i.e. helices that fail to fulfil this conditions
would exhibit the low-frequency tumbling, i.e., non-propulsive dynamics followed by the asynchronous tumbling for frequencies $\omega>\omega_\mathrm{I}$. Both regimes of tumbling motion take place \emph{in plane} of the field rotation (see Fig.~\ref{fig:lowfreq}), they are characterized by a single angular variable $\alpha$ and have been studied in detail, e.g. see Refs.~\cite{Pincus,Cebers1}.

\section*{Comparison to the experiment}

\begin{figure}[h] \centering
\includegraphics[width=0.4\textwidth]{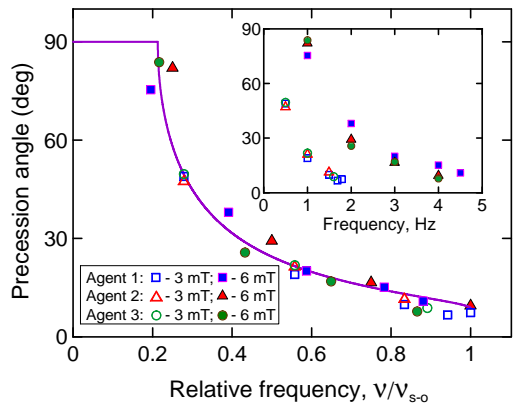}
\caption{
Precession angle as a function of the frequency ratio $\nu/\nu_{s-o}$. The symbols stand for the experimental results of \cite{Nelson} for three prototypes. The empty and filled symbols correspond to the applied field of 3 mT and 6 mT, respectively.  The solid line is the theoretical prediction by Eq.~(\ref{eq:sol24}) with parameter $\gamma=9.3$. The inset depicts the precession angle in \cite{Nelson} as a function of frequency $\nu$ (Hz).}
\label{fig:Nelson}
\end{figure}

Let us now compare the experimental results with our theoretical predictions. The experimental results \cite{Nelson} for the precession angle as a function
of the frequency of the rotating magnetic field are depicted in the inset of Fig.~\ref{fig:Nelson}. The data was obtained for three prototypes or `agents` (shown here as squares, triangles, and circles), having similar characteristics, i.e. microhelices with 3 full turns, helical radius $R=2.25$~$\:\mu$m, filament width $d=1.8$~$\mu$m, helical angle $\Theta=70^\circ$, fabricated from polymer composite with 2\% (vol) superparamagnetic magnetite $\sim$11~nm diameter nanoparticles. The empty and filled symbols correspond to
two different strengths of the applied external field, equal to 3 mT and 6 mT, respectively. Since the
magnetic and geometric properties of the helices are similar, their corresponding precession angles found at given field amplitude prove to be rather close.
However, the reported step-out frequencies $\nu_{s-o}=\omega_{\textrm{II}}/2\pi$ exhibit some scattering, especially noticeable at higher value of the magnetic field, $H=6$ mT, with $\nu_{s-o}=4$~Hz, $4.6$~Hz and $5.1$~Hz.

As follows from Eq.~(\ref{eq:sol24}), the precession
angle is a function of the steerability parameter $\gamma$ and the frequency ratio $\nu/\nu_{s-o}$. The re-scaled data shown in Fig.~\ref{fig:Nelson} falls on the master curve (\ref{eq:sol24}) with the single best-fitted parameter $\gamma=9.3$.
For $\gamma=9.3$ we find that $\nu_*/\nu_{s-o}=2\gamma/(1+\gamma^2)=0.21$. This prediction is also in an excellent agreement with the experimental values $\nu_*/\nu_{s-o}=0.20$, $0.22$ and
$0.25$, found in Ref.~\cite{Nelson} for the three prototypes.
Let us next estimate the limiting (minimal) value of the  precession angle, $\theta_{min}$,
corresponding to the step-out frequency $\nu_{s-o}$. Using Eq.~(\ref{eq:angle4}) it is $\theta_{min}=\arcsin{\sqrt{2/(1+\gamma^2)}}$.
For $\gamma=9.3$ we find $\theta_{min}=8.7^\circ$ which is quite close to the experimental measurement $\theta_{min}\approx 8^\circ$ \cite{Nelson}.

Now the angle $\Phi$ between the easy vector $\bm {n}$ and the axis $x_3$ of the helix corresponding to the best-fitted value of $\gamma=9.3$ can be determined.
From the definition of $\gamma$ we have $\Phi=\arctan{(\gamma/p)}$, where $p=\kappa_\perp/\kappa_\|$.
As done before in Ref.~\cite{ML}, we approximate the helix by the enclosing prolate spheroid
and use the explicit expressions for $\kappa_\perp$ and $\kappa_\|$ available for ellipsoidal particles (see Appendix B). This ratio depends only
on the aspect ratio $a/b$ of the spheroid. For a 3-turn helix this ratio is $a/b \approx 3 \pi/\tan \Theta$. For the helix angle $\Theta=70^\circ$, we find $a/b \simeq 3.4$. The aspect ratio can be alternatively estimated using the micrograph of the helix given in Fig.~2 in Ref.~\cite{Nelson}. The micrograph gives a slightly lower value, $a/b \approx 3$. Therefore, the corresponding values of the parameter $p=\kappa_\perp/\kappa_\|$ are 3.8 and 3.1, what finally determines the angle $\Phi$ of inclination of the easy-axis of magnetization $\bm {n}$ to the helical $x_3$-axis as $\Phi=68\div72^\circ$. These estimates are confirmed by the
rigorous particle-based calculations based on multipole expansion algorithm (see \cite{ML} for details). For example, for a helix with $\Theta=70^\circ$ ($a/b \simeq 2.61$) we found $p\simeq 3.07$ resulting in $\Phi\simeq 71.7^\circ$. For a slightly less slender helix with $\Theta=67^\circ$ (with $a/b \simeq 3.05$) we obtained $p\simeq 3.95$ yielding $\Phi\simeq 67^\circ$.

The propulsion velocity of the helical propeller along the axis of the field rotation, $U_Z$, can be determined from Eq.~(\ref{eq:u3}). Following the same arguments as in \cite{ML}, we assume helices with chirality along the $x_3$-axis, i.e., that in the body-fixed coordinate frame the only non-zero component of ${\boldsymbol {\mathcal {B}}}$ is $\mathcal{B}_{\|}$. Thus, in the low-frequency tumbling regime we have $U_Z=0$, whereas in the high-frequency regime it is $U_Z=-\omega c_{\theta}^2
\mathcal{B}_{\|}/\xi_{\|}$ (for details see \cite{ML}), where $\mathcal{B}_{\|}$ and $\xi_{\|}$ are the longitudinal (along the helix axes) components of the coupling and the translation viscous resistance tensors, respectively.
Substituting the value of precession angle from Eq.~(\ref{eq:sol24}) and normalizing the velocity with $R\omega_{s-o}$, with $R$ being the helix radius, yields
\begin{equation}
\frac {U_Z}{R\omega_{s-o}}= \mathrm{Ch} \frac {\omega}{\omega_{s-o}} \left[1-\frac {2}{1+\gamma^2}  \frac {\omega_{s-o}^2}{\omega^2} \left( 1+ {\sqrt{1-\frac {\omega^2}{\omega_{s-o}^2}}} \right) \right]\,, \label{eq:v}
\end{equation}
where $\mathrm{Ch}=- \mathcal{B}_{\|}/(\xi_{\|}R)$ is the chirality coefficient depending on the helix geometry only \cite{ML}.
Thus, similarly to the precession angle, the propulsion velocity is a
a function of the helix geometry (via parameters $\gamma$ and $\mathrm{Ch}$) and the frequency ratio $\nu/\nu_{s-o}$.
\begin{figure}[h] \centering
\includegraphics[width=0.4\textwidth]{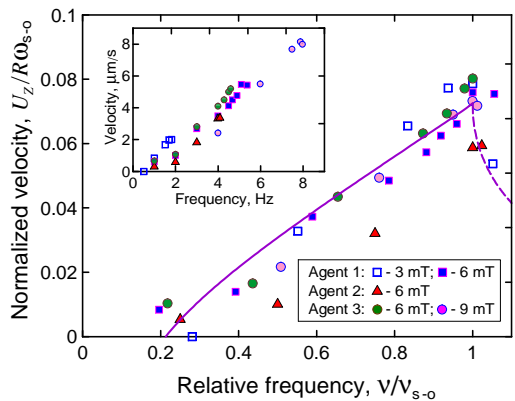}
\caption{
Scaled propulsion velocity vs. frequency of the rotating magnetic field normalized by the step-out frequency.
The experimental data (symbols) are from \cite{Nelson} for three different prototypes/agents.
The solid line is the theoretical prediction by Eq.~(\ref{eq:v}) with parameters $\gamma=9.3$ and $\mathrm{Ch}=0.075$. The dotted line stands for the velocity in the asynchronous regime \cite{ML}. The inset shows the experimental results in dimensional form.}
\label{fig:propulsion}
\end{figure}

In the inset to Fig.~\ref{fig:propulsion}, the experimental data \cite{Nelson} for the propulsion velocity as a function
of the frequency of the rotating magnetic field is shown.
The same symbol designations are used as in Fig.~\ref{fig:Nelson} and a new data for the
applied field strength of 9 mT was added.
As for the precession angle, the re-scaled data measured for three different values of the magnetic field collapse on the
the master curve (\ref{eq:v}). The previously best-fitted value of $\gamma=9.3$ and $\mathrm{Ch}=0.075$ are used. The data corresponding to the `agent' \#2 ($\triangle$) shows a considerable deviation from the master curve, what might be a result of a slightly lower value of $\mathrm{Ch}$.

To justify the parameter fitting, we computed $\mathrm{Ch}$ numerically
by the multipole expansion method described in \cite{ML}. Following \cite{Nelson}, we take the 3-turn helix with $\Theta=70^\circ$ and
$R/r=2.5$ (corresponding to $a/b\simeq 2.61$), where $r$ stands for the filament radius. For such geometry we find $\mathrm{Ch}=0.056$.
However for a slightly more slender helix with $\Theta=67^\circ$ (corresponding to $a/b\approx 3.05$, which agrees with the micrograph in Fig.~2 in \cite{Nelson}), we obtain $\mathrm{Ch}=0.072$ in a very good agreement with the best-fitted value of $0.075$.

\section*{Magnetic properties of helices}

In this section we study the magnetic properties of helices. We assume that helices are superparamagnetic:
they do not possess remanent spontaneous magnetization and become magnetized only in an external magnetic field.
In practice, such helices are microfabricated by the solidification (photopolymerization) of a polymer matrix comprising the embedded
superparamagnetic nanoparticles of typical size $\sim 10$~nm \cite{Nelson,Suter1}. The matrix crosslinking
can take place both in the absence of external magnetic field \cite{Suter1,Suter0,Suter2} as well as with applied static homogeneous magnetic field
\cite{Peters}. In the former case the magnetic particles are randomly distributed (reference helices), whereas in the later case there is
an anisotropic particle distribution (helices with adjusted anisotropy). Here we investigate the case of
helices with \emph{random} spatial distribution of superparamagnetic inclusions. Therefore, the magnetic susceptibility
of the helix bulk material is an isotropic property characterized by the scalar parameter $\chi_0$ determined
by the values of concentration and magnetic moments of nanoparticles \cite{Rosen}.

Our aim is determining the \emph{effective} magnetic susceptibility $\bm \chi$ of the helix as \emph{a whole}.
We define $\bm \chi$ as the coefficient of proportionality between the helix
magnetization $\bm M={\boldsymbol {\mathcal {M}}}/V$
(${\boldsymbol {\mathcal {M}}}$ is magnetic moment of helix acquired in the external field magnetic field $\bm H$ and $V$ is the helix volume) and the value of this field:
${\bm {M}}={\boldsymbol {\chi}}\cdot{\bm H}$. Let us comment on this relation. Typically, magnetic susceptibility is defined as a coefficient
of proportionality between magnetization $\bm M$ and \emph{internal} magnetic field $\bm H_{in}$. Internal magnetic field is a sum
of the external field, ${\bm H}$, and the demagnetizing magnetic field, ${\bm H}_d$, owing to the magnetic
material itself, $\bm H_{in}={\bm H}+{\bm H}_d$ \cite{LL8}. For  magnetic particles the
demagnetizing field ${\bm H}_d$ depends on the body geometry, whereas the susceptibility $\bm \chi$
is a property of a magnetic material only, i.e. geometry independent.
For the helical geometry, the internal field $\bm H_{in}$ proves to be fundamentally inhomogeneous
one, and, therefore it is advantageous to characterize the apparent, or, effective susceptibility of the helix as a whole
with geometry dependent tensor $\bm \chi$. For sufficiently slender helices one can estimate the effective susceptibility tensor $\bm \chi$ as
a function of magnetic susceptibility $\chi_0$ of the helix material and its geometry in the framework of slender body (SB) approximation.

\subsection*{Slender body approximation}

The SB approximation assumes that locally a helical filament can be considered as a thin straight cylinder. We study the case of an \emph{elliptical} cross-section of the filament
with semi-axis $\mathtt{a}$ having a fixed component along the helical axis and semi-axis $\mathtt{b}$ oriented normally to the helical axis. Assumption of slenderness applies to helices with typical dimensions, i.e. the radius $R$ and the pitch $P$, satisfying $R, \,P \gg \max{(\mathtt{a},\mathtt{b})}$. In the experiments the helices are typically not slender (e.g. \cite{Nelson,Suter2,Peters}), however,
SB approximation allows the derivation of the closed-form formulae for the effective susceptibility and explains qualitatively the experimentally
observed phenomena.

Here we consider two types of helices: \emph{normal} helices with the cross-section elongated in the direction transverse to the helical axis ($\mathtt{b}>\mathtt{a}$),
and \emph{binormal} helices with longer cross-sectional axis having a fixed component along the helical axis ($\mathtt{a}>\mathtt{b}$).
In what follows, we shall consider the normal helix.

In the body-fixed coordinate frame $x_1x_2x_3$ with the helix axis oriented along $x_3$,
the equation for the helix centerline can be written in the following parametric representation \cite{Fonseca}
\begin{equation}
\textbf{X}(s)=\left[\frac {\kappa}{\lambda^2}\cos (\lambda s),\frac {\kappa}{\lambda^2}\sin (\lambda s),\frac {\tau}{\lambda} s\right]\,.
\label{eq:par5}
\end{equation}
Here $s$ is the arc length and $\lambda={1}/{\sqrt {R^2+\frac {P^2}{4\pi^2}}}$.
Curvature $\kappa$ and torsion $\tau$ are defined via helix radius $R$ and pitch $P$ as $\kappa=R\lambda^2$,
$\tau=\frac {P}{2\pi}\lambda^2$.

Let $\{\bm {d}_1, \bm {d}_2, \bm {d}_3\}$ be the right-handed director basis defined at
each position $s$ along the axis of the filament \cite{Fonseca}:
\begin{eqnarray}
\bm {d}_1(s)&=&\left[\frac {\tau}{\lambda}\sin (\lambda s),-\frac {\tau}{\lambda}\cos (\lambda s),\frac {\kappa}{\lambda} \right]\,, \nonumber \\
\bm {d}_2(s)&=&[\cos (\lambda s),\sin (\lambda s),0]\,, \label{eq:d} \\
\bm {d}_3(s)&=&\left[-\frac {\kappa}{\lambda}\sin (\lambda s),\frac {\kappa}{\lambda}\cos (\lambda s),\frac {\tau}{\lambda} \right]\,.\nonumber
\end{eqnarray}
$\bm {d}_3=\partial \textbf{X}(s)/\partial s$ is the vector tangent to
the centreline of the filament.  Vectors $\bm {d}_1$ (binormal) and $\bm {d}_2$ (normal) are assumed to be parallel, correspondingly,
to the semi-axes of the filament cross-section (e.g. for the normal helix $\bm {d}_1$ and $\bm {d}_2$ are parallel to the short and the long semi-axes, respectively).

The magnetic susceptibilities of a cylinder along the three principal axes read
\cite{LL8}
\begin{equation}
\chi_{1}=\frac {\chi_0}{1+4\pi\chi_0N_1}\,,\,\chi_{2}=\frac {\chi_0}{1+4\pi\chi_0N_2}
\,,\,\chi_{3}=\chi_0\,.
\label{eq:cyl}
\end{equation}
Here $N_1$ and $N_2=1-N_1$ are the demagnetizing factors along the axis  $\bm {d}_1$ and $\bm {d}_2$,
respectively, and we also assumed the zero value of the demagnetizing  factor
along the long axis of cylinder, $N_3=0$. The explicit expressions for demagnetizing factors are given in the Appendix C.

Eq.~\ref{eq:cyl} indicates that in the applied magnetic field helical segments are polarized differently along the major axes:
the easy direction is along the centerline, and the hard one is along the shorter cross-section. This property
leads to the apparent anisotropy of magnetic susceptibility $\bm \chi$ of the helix as a whole.

The SB approximation is the \emph{local} theory: the magnetization of each segment is determined only by its geometry and by
the applied magnetic field (see Eq.~(\ref{eq:chi})). In other words, in the framework of SB approximation different parts of the helix do not interact, i.e. magnetize
independently from each other. Therefore, the effective susceptibility $\bm \chi$ of helix proves to be an additive
property and can be determined by integration
\begin{equation}
\bm \chi=L^{-1}\int_0^L \left(\chi_{1}{\bm d}_1 {\bm d}_1+\chi_{2}{\bm d}_2 {\bm d}_2+\chi_{3}{\bm d}_3 {\bm d}_3\right)ds\,,
\label{eq:chi1}
\end{equation}
where $L$ is the helix length along the centerline.

Let us denote the eigenvalues of matrix $\bm \chi$ in the ascending order as $\chi_I\le \chi_{II}\le \chi_{III}$.
Generally, all three eigenvalues are different and helix possesses bi-axial magnetization. However, for the \emph{integer} number of turns in the framework of SB approximation, the susceptibility tensor becomes \emph{uniaxial}: two out of three eigenvalues coincide. In present study we shall consider this simple and practically relevant situation. The direct integration in Eq.~(\ref{eq:chi1}) using Eqs.~(\ref{eq:cyl})-(\ref{eq:d}) demonstrates that the eigenvectors of $\bm \chi$ are aligned with the BCS axes $x_1x_2x_3$; the magnetic anisotropy parameter $\Delta \chi$, defined as the difference of eigenvalues along the anisotropy (i.e. helical) axes and in transverse direction, $\Delta \chi=\chi_{\|}-\chi_{\perp}$, reads
\begin{equation}
\Delta \chi=\textstyle \frac 12 [(\chi_{3}-\chi_{1})(3\cos ^2 \Theta-1)+\chi_{1}-\chi_{2}]\,.
\label{eq:del}
\end{equation}
Particularly simple form of magnetic anisotropy parameter can be obtained for the case when $4\pi\chi_0 \ll 1$.
The polymer composite used for nanomotor fabrication in Ref.~\cite{Suter1,Suter0,Suter2} satisfies this condition. Indeed, taking the volume fraction of superparamagnetic inclusions $\phi=0.02$, their mean diameter $d_{p}=11$~nm \cite{Suter1} and the saturation magnetization of the magnetite $M_s=281$~Gs \cite{Suter2}, at $T=300$~K we can estimate that $4\pi \chi_0=(2\pi^2/9)\phi M_s^2 d_{p}^3/(k_B T)\approx 0.1$ \cite{Rosen}. Then taking the asymptotic small-$\chi_0$ limit of susceptibilities in Eq.~(\ref{eq:cyl}), $\Delta\chi$ in Eq.~(\ref{eq:del}) can be further simplified into
\begin{equation}
\frac {\Delta \chi}{2\pi\chi_0^2}=N_1(3\cos ^2 \Theta-1)+1-2N_1\,.
\label{eq:del1}
\end{equation}
The analogous result for the binormal helix is readily obtained from Eqs.~(\ref{eq:del}) and (\ref{eq:del1})
by interchanging indices $1\leftrightarrow 2$.

The obtained result (\ref{eq:del1}) indicates that magnetically, integer-number-of-turns helix is
equivalent to a polarized spheroid with its polarization/easy-axis aligned along the helical $x_3$-axis. The slender helices with
a small pitch angle $\Theta<\Theta_*$ are characterized by the positive value of the anisotropy parameter $\Delta \chi$ (i.e. equivalent to prolate spheroid), whereas for the tight helices with
high values of $\Theta>\Theta_*$ the anisotropy parameter becomes negative (i.e. equivalent to oblate spheroid or disk). The critical helix angle $\Theta_*$ at which $\Delta\chi$ changes sign
is found from the relation $\cos^2{\Theta_*}=1-1/(3N_1)$.
\begin{figure}[h] \centering
\includegraphics[width=0.45\textwidth]{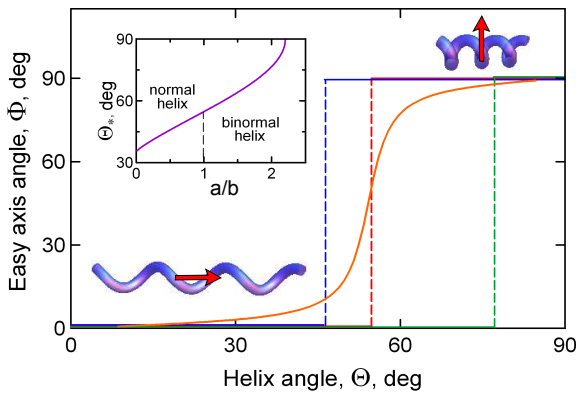}
\caption{
Inclination angle $\Phi$ of the easy-axis as a function of the helix angle $\Theta$ for regular (red),
normal (blue) and binormal (green) helices with integer number of turns. The
aspect ratio of the filament cross-section for normal and binormal helices is $\mathtt{a}/\mathtt{b}=2$. The orange curve corresponding to a regular helix with $3.3$ full turns
illustrates the effect of imperfectness on $\Phi$ (see details in the text); the helices and corresponding arrows illustrate the orientation
of the easy-axis depending on the pitch angle $\Theta$. The inset depicts the critical helix angle $\Theta_*$ vs. the filament cross-section aspect ratio $\mathtt{a}/\mathtt{b}$.
\label{fig:angle}}
\end{figure}
It is depicted in the inset to Fig.~\ref{fig:angle} as a function of the aspect ratio $\mathtt{a}/\mathtt{b}$ of
the filament cross-section. For regular helix with a circular cross-section ($\mathtt{a}=\mathtt{b}$) we have $\Theta_*^{r}=54.7^\circ$. The
values of the critical angle $ \Theta_*$ for the normal and binormal helices prove to be \emph{strongly asymmetric}
relatively to its value for regular helix. For example, for $\mathtt{a}/\mathtt{b}=2$,  $\Theta_*^{n}=45.8^\circ$
and $ \Theta_*^{bn}=76.6^\circ$ for the normal and binormal helices, respectively. The minimal
value of this critical angle for the normal helix with infinitely thin filament cross-section ($a/b \rightarrow 0$) is $\Theta_{*}^{n}=35.3^\circ$, whereas for the binormal helix it
reaches it maximal value $\Theta_{*}^{bn}=90^\circ$ already for aspect ratio $a/b\simeq 2.2$.
The angle $\Phi$ formed by the main eigenvector (corresponding to the maximal eigenvalue $\chi_{III}$) of the
susceptibility tensor and the helix axes is depicted in Fig.~\ref{fig:angle} as a function of the helix angle $\Theta$ for regular (red), normal (blue) and binormal (green) helices.
As seen, all dependencies are \emph{step-like} functions: $\Phi={\mathcal H}(\Theta-\Theta_*)$, where ${\mathcal H}(x)$
is the Heaviside function. This idealized solution is due to three simplifying assumptions:
(i) local magnetization, (ii) integer number of helical turns and (iii) homogeneity of the
geometrical and magnetic properties of helices. Any violation of (i)--(iii) or \emph{imperfectness} should lead to deviation from the ideal dependence and
to smoothing out of the step-like profile as illustrated in Fig.~\ref{fig:angle} by the orange line corresponding to a helix with non-integer number of turns. The detailed analysis of magnetization of polarizable helices requires numerical computations that are beyond the scope of the present study and will be considered elsewhere.

The idealized SB approximation allows to understand qualitatively the difference in the dynamics
of slender ($\Theta < \Theta_*$) and tight ($\Theta >\Theta_*$) helices. In the former case, the easy-axis inclination angle $\Phi=0$ so that the condition $p\tan \Phi > 1$ (obligatory for transition to wobbling, see the details above)
is violated. That means that helices with pitch angles $\Theta < \Theta_{*}$ would only tumble and not propel. In contrast, for $\Theta > \Theta_{*}$, the inclination angle is at maximum, $\Phi=90^\circ$, i.e. the optimal orientation for propulsion. These findings are in accord with recent experimental observations in \cite{Nelson}, where helices with with pitch angles $\Theta=40^\circ$, $50^\circ$ and $60^\circ$ showed tumbling
for any driving frequencies, while only the helices with the pitch angle $\Theta=70^\circ$ exhibited corkscrew-like propulsion. We should mention, however, that within the SB approximation framework, the anisotropy is always uniaxial, meaning that tight helices possess easy-plane of magnetization (disk-like polarization with $\Delta\chi<0$), while effective propulsion requires, besides alignment along the axis of field rotation (i.e. $\Phi\sim 90^\circ$) also magnetic anisotropy in the transverse plane. We anticipate that aforementioned imperfectness should yield deviation from the uniaxial anisotropy, e.g. two distinct non-zero anisotropy parameters, ${\Delta\chi}_1=\chi_{III}-\chi_{II}$ and ${\Delta\chi}_2={\chi}_{II}-\chi_{I}$.

\section*{Discussion and concluding remarks}

We developed the theory for dynamics and polarization of superparamagnetic chiral nanomotors powered by a rotating magnetic field. Depending on their geometry,  magnetic properties and the parameters  of actuating magnetic field (i.e. frequency and amplitude), the nanomotors are involved into synchronous motion (tumbling or wobbling) or twirl asynchronously. The effective nanomotor propulsion is enabled as a combined effect of two different factors. First factor is the ``steerability" of the nanomotor, i.e. its ability to undergo synchronous precessive motion and propulsion. Mathematically, steerability can be characterized by the parameter $\gamma=p \tan \Phi$ that depends on both geometric and magnetic properties of the nanomotor.  The nanomotors with $\gamma<1$ are not propulsive and undergo tumbling for all driving frequencies, while $\gamma>1$ controls the critical frequency of tumbling-to-wobbling transition, $\omega_*=(A/2) \sin{2\Phi}$ and the minimal value of the wobbling angle $\theta_{min}\sim \gamma^{-1}$ (see Eq.~ \ref{eq:angle4}). Increasing the slenderness of the propeller (i.e. increasing $p=\kappa_\perp/\kappa_\|$) and/or the inclination of the easy-axis of magnetic anisotropy relatively to the helix axis $\Phi$, results in narrowing of the interval of tumbling frequencies, $[0,\omega_*]$ and better alignment via lowering of $\theta_{min}$, i.e. improved steerability. The maximal steerability is attained as $\gamma \rightarrow \infty$ when the easy-axis is oriented transverse to the helix axis, i.e. $\Phi\approx90^\circ$. In this case the propulsion is tumbling- and wobbling-free as the nanomotor aligns parallel to the axis of the field rotation for all frequencies. This situation corresponds to, e.g., nanohelices with ``adjusted" easy-axis anisotropy \cite{Peters}.

The second factor is the anisotropy parameter of the effective susceptibility in the plane of field rotation,  $\Delta \chi_\perp$. This anisotropy parameter defines the maximal value of the step-out frequency $\omega_{s-o}=\Delta \chi_\perp H^2 V/2\kappa_\|$ at the best possible orientation, $\Phi \rightarrow \pi/2$ (see Eq.~\ref{eq:so}). In the synchronous wobbling regime the propulsion is \emph{geometric}, as as the swimming speed $U_Z\approx \mathrm{Ch}\, R\, \omega$ for sufficiently large $\gamma$ (see Eq.~(\ref{eq:v})), i.e. it is the same for all values of $\Delta\chi_\perp \ne 0$. The step-out frequency, however, determines the upper limit for the propulsion speed,   ${U_Z}_{max}=\mathrm{Ch} \,\omega_{s-o}\,R$.

The predictions of our theory for the dynamic orientation and propulsion speed are in excellent agreement with available experimental results (see Figs.~\ref{fig:Nelson}--\ref{fig:propulsion}). Note that $\gamma$ could not be estimated directly from the experiments in \cite{Nelson} since the easy-axis orientation, i.e. $\Phi$, was not reported, and it was treated as fitting parameter. The chirality coefficient, $\mathrm {Ch}$, determined self-consistently agrees with the corresponding estimated experimental value.

The developed slender body (SB) theory provides a qualitative description of the effective polarization of the nanomotors. In particular, it predicts the uniaxial magnetic anisotropy of helices with integer number of turns. Orientation of the easy-axis is a step-function of the helix pitch angle, i.e. slender helices possess an easy-axis aligned along the helical axis ($\Phi=0$, $\Delta \chi>0$), while tight helices possess (disk-like) easy-plane anisotropy ($\Phi=90^\circ$, $\Delta \chi<0$). Thus, within the SB approximation framework slender helices are not steerable ($\gamma=0$), while tight helices are shown to be optimally oriented for propulsion by the rotating field in agreement with experimental observations \cite{Nelson}. However, it is impossible to estimate the propulsion velocity of the tight helices in the SB  framework, as $\Delta\chi_\perp=0$ due to magnetization isotropy in the transverse plane. In practice, however, transverse magnetization of nanomotors with non-adjusted easy-axis is anisotropic owing to potential shape effects, non-slenderness, fluctuations in the spatial distribution of superparamagnetic inclusions, etc. In \cite{Peters} non-adjusted helices exhibited nearly optimal orientation, $\Phi \approx 90^\circ$, while the propulsion velocity of adjusted and non-adjusted helices under similar conditions felt on the same straight line when plotted vs. driving frequency in accord with our arguments above. The difference in the step-out frequency, $\nu_{s-o}\approx 5$~Hz and $\nu_{s-o}\approx 18$~Hz for non-adjusted and adjusted helices \cite{Peters}, respectively, indicates that  $\Delta\chi^{adjusted}_\perp/\Delta\chi^{non-adjusted}_\perp \approx 3.6$. The detailed theoretical study of the apparent polarization of superparamagnetic helical nanomotors will be a subject of our future work.

\section*{Acknowledgement}

The authors would like to thank Christian Peters and Kathrin Peyer for providing details of their experiments. This work was partially supported by the Japan Technion Society
Research Fund (A.M.L.) and by the Israel Ministry for Immigrant Absorption (K.I.M.).


\begin{thebibliography}{99}
\bibitem{note1} Generally, the principal rotation axis of a helix slightly deviates from $x_3$, however,
this misalignment has a negligible effect on the dynamics and can be ignored \cite{ML}.
\end{thebibliography}

\section*{Appendix A: Rotation matrix}


We use the definition of the three Euler angles $\varphi$, $\theta$ and $\psi$
following Ref.~\cite{LL}. The components of any vector $\textbf{W}$ in the
body-fixed coordinate system (BCS) and in the laboratory coordinate system (LCS)
are  determined from the relation $ \textbf{W}^{BCS}=\textbf{R}\cdot\textbf{W}$,
where $\bf {R}$ is the rotation matrix. The rotation matrix is expressed explicitly
via the Euler angles \cite{D}
\[\bf {R} = \begin{pmatrix}
c_{\varphi}c_{\psi}-s_{\varphi}s_{\psi}c_{\theta} & s_{\varphi}c_{\psi}+c_{\varphi}s_{\psi}c_{\theta} &  s_{\psi}s_{\theta} \\
-c_{\varphi}s_{\psi}-s_{\varphi}c_{\psi}c_{\theta} & -s_{\varphi}s_{\psi}+c_{\varphi}c_{\psi}c_{\theta} &  c_{\psi}s_{\theta} \\
s_{\varphi}s_{\theta} & -c_{\varphi}s_{\theta} &  c_{\theta}
\end{pmatrix}\,,
\]
where we use the compact notation,  $s_{\psi}=\sin{\psi}$, $c_{\theta}=\cos{\theta}$, etc.

\section*{Appendix B: Rotational viscous resistance coefficients of a helix}

We approximate the rotational viscous resistance coefficients of a helix by the corresponding values for a spheroid enclosing the helix. Let $a$ and $b$ be,
correspondingly, the longitudinal (along the symmetry axis) and transversal semi-axes of the spheroid. The respective viscous resistances due to rotation around the symmetry axis and
in perpendicular direction read \cite{Jeffrey}
\begin{equation}
\kappa_{\|}=2\eta V n_{\perp}^{-1}\,,\,\,\,\,\,\,\kappa_{\perp}=2\eta V \frac {a^2+b^2}{a^2 n_{\|}+b^2n_{\perp}}\,,
\label{eq:fric}
\end{equation}
where $\eta$ is the dynamic viscosity of the liquid, $V$ is the spheroid volume, $n_{\|}$ and
$n_{\perp}=(1-n_{\|})/2$ are the depolarizing factors of the spheroid.
For the prolate spheroid with $a>b$ and eccentricity $e=\sqrt{1-b^2/a^2}$ the depolarizing factor along the
symmetry axis reads \cite{LL8}
\begin{equation}
n_{\|}=\frac {1-e^2}{e^3}\left( \frac 12 \ln \frac {1+e}{1-e}-e\right)\,.
\label{eq:fric2}
\end{equation}

\section*{Appendix C: Demagnetizing factors of infinitely long elliptic cylinder}

The demagnetizing factor $N$ of infinitely long elliptic cylinder with the cross-section with corresponding semi-axes $\mathtt{a}$ and $\mathtt{b}$
has been found in Ref.~\cite{Brown}:
\begin{eqnarray}
N&=&(2\pi)^{-1}\left[4\arctan \frac {\mathtt{a}}{\mathtt{b}} + \frac {2\mathtt{b}}{\mathtt{a}} \ln \frac {\mathtt{b}}{\mathtt{a}}\, + \right.\nonumber \\
&& \quad \left. \left( \frac {\mathtt{a}}{\mathtt{b}} - \frac {\mathtt{b}}{\mathtt{a}} \right)\ln \left(1+ \frac {\mathtt{b}^2}{\mathtt{a}^2} \right) \right]\,.\nonumber
\end{eqnarray}
At $\mathtt{a}>\mathtt{b}$ it determines the demagnetization factor $N_1$ along the short axes. The demagnetizing
factor $N_2$ can be found either by the permutation $\mathtt{a}\leftrightarrow \mathtt{b}$ or from the equality $N_1+N_2=1$.
For the regular helix with circular cross-section $N_1=N_2=1/2$.



\end{document}